\documentclass[runningheads]{llncs}
\usepackage{graphicx}
\usepackage{mathtools}
\usepackage[misc,geometry]{ifsym} 

\usepackage{tabularx}
\usepackage{multirow}
\newcommand{\etal}{\textit{et al}. }
\newcolumntype{L}[1]{>{\raggedright\arraybackslash}p{#1}}
\newcolumntype{C}[1]{>{\centering\arraybackslash}p{#1}}
\newcolumntype{R}[1]{>{\raggedleft\arraybackslash}p{#1}}

\begin{document}
\title{Anatomy-Aware Cardiac Motion Estimation}
\titlerunning{AATracker}
\authorrunning{P. Chen et al.}
\author{Pingjun Chen \inst{1} \orcidID{0000-0003-0528-1713} \thanks{This work was carried out during the internship of the author at United Imaging
Intelligence, Cambridge, MA 02140} \and
Xiao Chen\inst{2} \orcidID{0000-0002-7147-7311} \and
Eric Z. Chen\inst{2} \orcidID{0000-0001-5002-720X} \and
Hanchao Yu \inst{3}\and 
Terrence Chen \inst{2} \and
Shanhui Sun \inst{2} \orcidID{0000-0001-9841-8592} \Letter 
}
\institute{University of Florida, Gainesville, FL 32611, USA \and
United Imaging Intelligence, Cambridge, MA 02140, USA \and
University of Illinois at Urbana-Champaign, Urbana, IL 61801, USA\\
Corresponding author: \email{shanhui.sun@united-imaging.com}}
\maketitle

\begin{abstract}
Cardiac motion estimation is critical to the assessment of cardiac function. Myocardium feature tracking (FT) can directly estimate cardiac motion from cine MRI, which requires no special scanning procedure. However, current deep learning-based FT methods may result in unrealistic myocardium shapes since the learning is solely guided by image intensities without considering anatomy. On the other hand, motion estimation through learning is challenging because ground-truth motion fields are almost impossible to obtain. In this study, we propose a novel Anatomy-Aware Tracker (AATracker) for cardiac motion estimation that preserves anatomy by weak supervision. A convolutional variational autoencoder (VAE) is trained to encapsulate realistic myocardium shapes. A baseline dense motion tracker is trained to approximate the motion fields and then refined to estimate anatomy-aware motion fields under the weak supervision from the VAE. We evaluate the proposed method on long-axis cardiac cine MRI, which has more complex myocardium appearances and motions than short-axis. Compared with other methods, AATracker significantly improves the tracking performance and provides visually more realistic tracking results, demonstrating the effectiveness of the proposed weakly-supervision scheme in cardiac motion estimation.

\end{abstract}

\section{Introduction}
\begin{figure}[t]
\includegraphics[width=\textwidth]{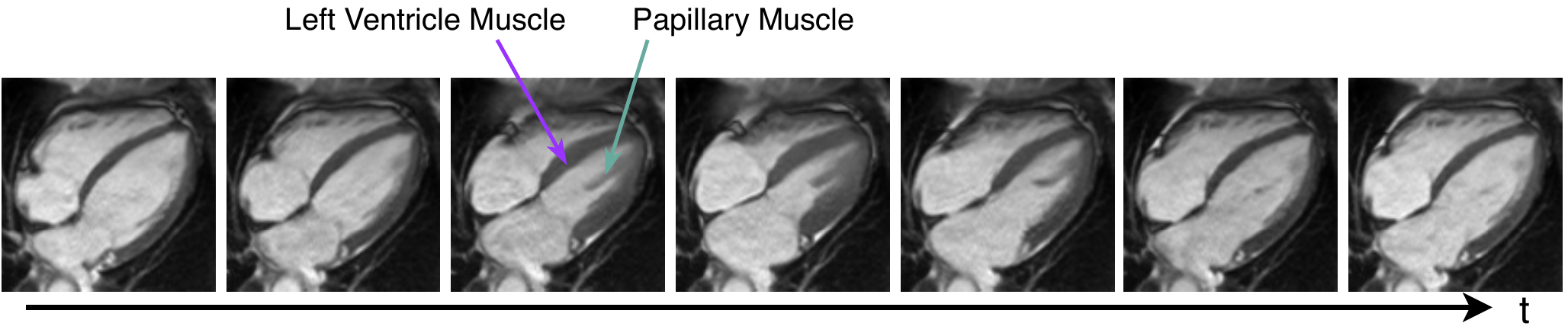}
\caption{Full cardiac cycle in Cine MRI. Example frames from a long-axis cine MRI show the heart motion starting from relaxation to contraction and then back to relaxation. The left ventricular (LV) muscle (V-shape dark region) undergoes large deformation during the cycle. Papillary muscle appears similar to the LV muscle.}
\label{fig:cineMRI_cardiac_cycle}
\end{figure}

Accurate cardiac motion estimation plays a critical role in cardiac function assessment, such as myocardium strain, torsion, and dyssynchrony, which have been demonstrated as sensitive and early indicators of myocardial disorders \cite{muser2018clinical}. Myocardial feature tracking (FT) can provide motion estimation from breath-hold 2D cine MRI, which is recommended by the American Heart Association (AHA) for clinical routine \cite{writing2010accf}. Cardiac MRI feature tracking (CMR-FT) \cite{puyol2018fully,heinke2019towards} estimates time-varying cardiac motion from cine MRI that usually includes a complete cycle of cardiac contraction and relaxation, as shown in Fig. \ref{fig:cineMRI_cardiac_cycle}. Starting from an initial image frame (usually end-diastole ED) as a reference, the next frame in time as a source image is compared, and motion occurred in-between is estimated. Then the source image becomes a new reference frame, and the process is repeated for all the consecutive frames to obtain the motion for the full cardiac cycle. 

CMR-FT is a challenging topic still under active investigation, as addressed e.g. in  \cite{puyol2018fully,vigneault2017feature,krebs2019learning,qin2018joint,zheng2019explainable,yu2020foal,yu2020motion}. A conventional image registration based method was proposed in \cite{puyol2018fully} to estimate left ventricle motion using 2D B-spline free form deformation (FFD). Vigneault~\etal in \cite{vigneault2017feature} proposed to perform cardiac segmentation and then apply B-Spline registration on the boundary points to track myocardium. This approach requires additional segmentation work. Tracking only boundary points also limits the motion estimation accuracy since both image features inside and outside the myocardium are not considered. Recently, Krebs~\etal in \cite{krebs2019learning} presented a variational autoencoder based image registration algorithm for estimating motion field for two consecutive frames from cine CMR. The VAE encoder takes the two images and encode them into latent variables which generate cardiac deformations via decoder. Because myocardium has similar image appearances with neighboring tissues/organs such as the papillary muscle, mere image-based motion estimation will face severe ambiguity in these regions, which causes anatomically unrealistic tracked results. Qin~\etal proposed to jointly learn motion estimation and segmentation using a supervised deep learning method \cite{qin2018joint}.  Zheng~\etal introduced a semi-supervised learning method for apparent flow estimation aiming for explainable cardiac pathology classification \cite{zheng2019explainable}. Both studies make use of raw MR images and cardiac segmentation together for accurate flow estimation. 

\begin{figure}[bt]
\includegraphics[width=\textwidth]{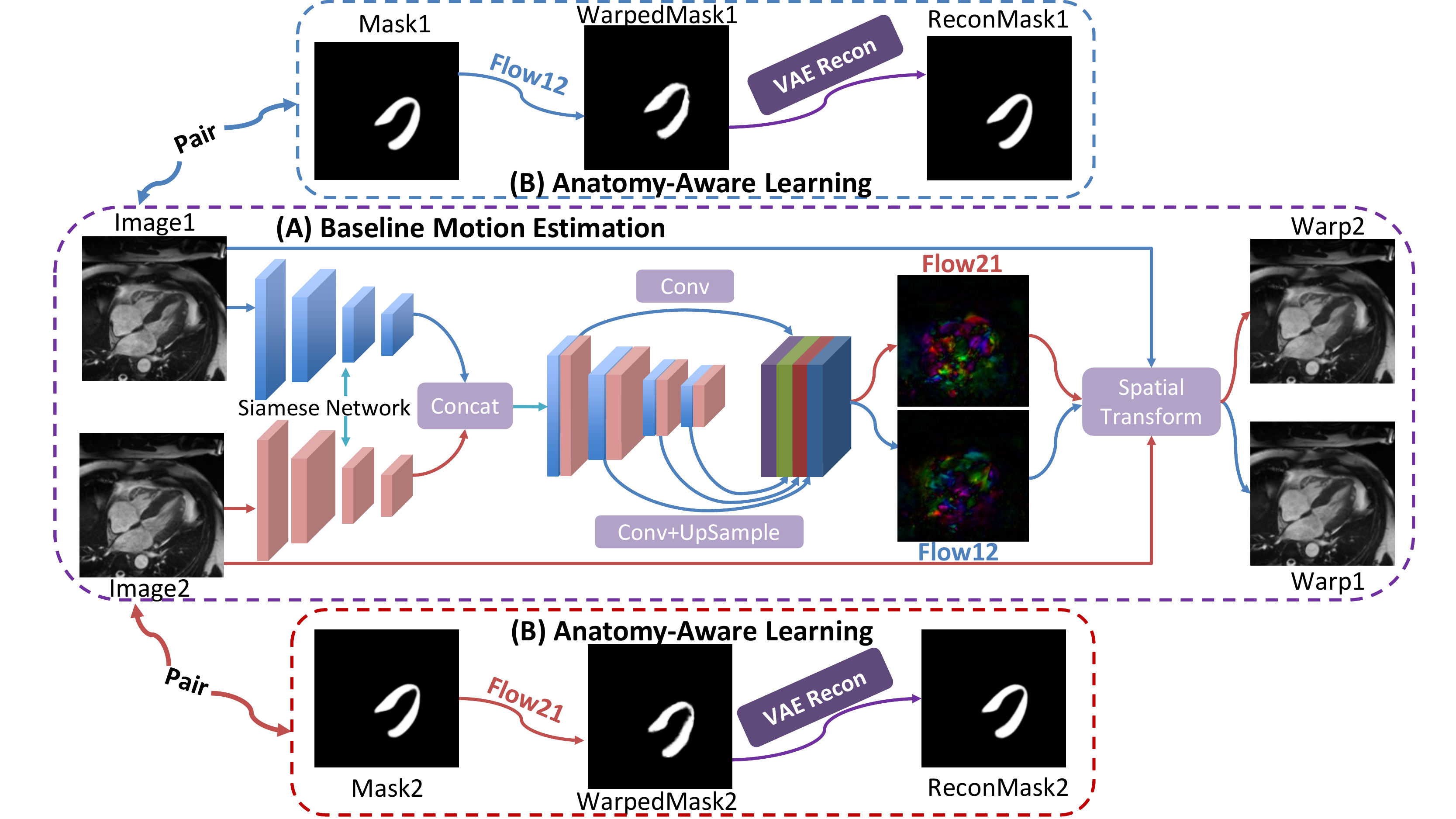}
\caption{The illustration of the baseline unsupervised motion estimation model and the weakly-supervised anatomy-aware model (AATracker). Module (A) presents the baseline, and Module (B) adds shape constraints to enhance the baseline motion estimation. Besides the image intensity loss and motion field regularization in the baseline model, loss between the warped mask with the VAE refined results (anatomy loss) and loss between the warped mask with its VAE reconstruction (reconstruction loss) are introduced in Module (B) to constrain the motion estimation for anatomy awareness.}
\label{fig:motion_estimation}
\end{figure}

In this study, we propose an end-to-end framework to estimate cardiac motion that is aware of the underlying anatomy through shape-constraints in a weakly supervision manner, coined as anatomy-aware tracker (AATracker). We first train an unsupervised CNN-based dense tracker as the baseline. We then train a convolutional VAE model that learns the latent space of realistic myocardium shapes. We apply the trained VAE model to baseline-tracked myocardium and treat the anatomically reasonable myocardium masks from the VAE as self-learned shape constraints. The baseline model is further refined as AATracker using shape constraints for anatomy awareness.  We evaluate the proposed method on the Kaggle Cardiovascular Disease dataset\footnote{https://www.kaggle.com/c/second-annual-data-science-bowl/data}.

The main contributions of this work are as follows: (1) We present AATtracker, an end-to-end anatomy-aware cardiac motion estimation model via weak supervision. (2) We employ VAE to constrain motion estimation for anatomy awareness. (3) AATtracker significantly improves the performance of myocardium feature tracking compared to baseline and conventional methods.

\section{Method}
\subsection{Unsupervised Motion Estimation}
The network structure for baseline cardiac motion estimation is shown in the middle part of Fig. \ref{fig:motion_estimation}. Two images, source and target at different time points of cine MRI, are inputs to the network. Three main modules include a Siamese network for mutual image feature extraction, a  multi-scale decoder for flow field generation, and a spatial transform that warps the source image with the flow fields \cite{bertinetto2016fully,jaderberg2015spatial}. Unlike previous studies \cite{balakrishnan2018unsupervised,mansilla2020learning}, the motion estimation framework here is symmetrical, inspired by traditional symmetric registration algorithms \cite{avants2008symmetric,wu2014s}. 

For an input image pair ($\mathcal{I}_{1}$, $\mathcal{I}_{2}$), $\mathcal{F}_{12}$ is the flow field from $\mathcal{I}_{1}$ to $\mathcal{I}_{2}$, and $\mathcal{F}_{21}$ is the reverse. Using $\bigotimes$ as the warping operator, $\mathcal{I}_{1}^{'}=\mathcal{F}_{12}\bigotimes \mathcal{I}_{1}$ and $\mathcal{I}_{2}^{'} = \mathcal{F}_{21} \bigotimes \mathcal{I}_{2}$ are the warped results of $\mathcal{I}_{1}$ and $\mathcal{I}_{2}$ via the spatial transform, respectively. The loss function enforcing warping consistency is defined as $\mathcal{L}_{cons} = \Vert \mathcal{I}_{1} - \mathcal{I}_{2}^{'} \Vert + \Vert \mathcal{I}_{2} - \mathcal{I}_{1}^{'}\Vert$. We add Huber loss $\mathcal{L}_{H} = \mathcal{H}(\mathcal{F}_{12})+\mathcal{H}(\mathcal{F}_{21})$ on the flow fields as the regularizer for motion smoothness  \cite{huber1992robust}. The loss function for the baseline model is then formulated as:
\begin{equation}
    \mathcal{L}_{base} = \mathcal{L}_{cons} + \lambda_{H}\mathcal{L}_{H},
\end{equation}
where $\lambda_{H}$ is the Huber loss weight.

\subsection{Myocardium Feature Tracking}
\begin{figure}[bt]
\includegraphics[width=\textwidth]{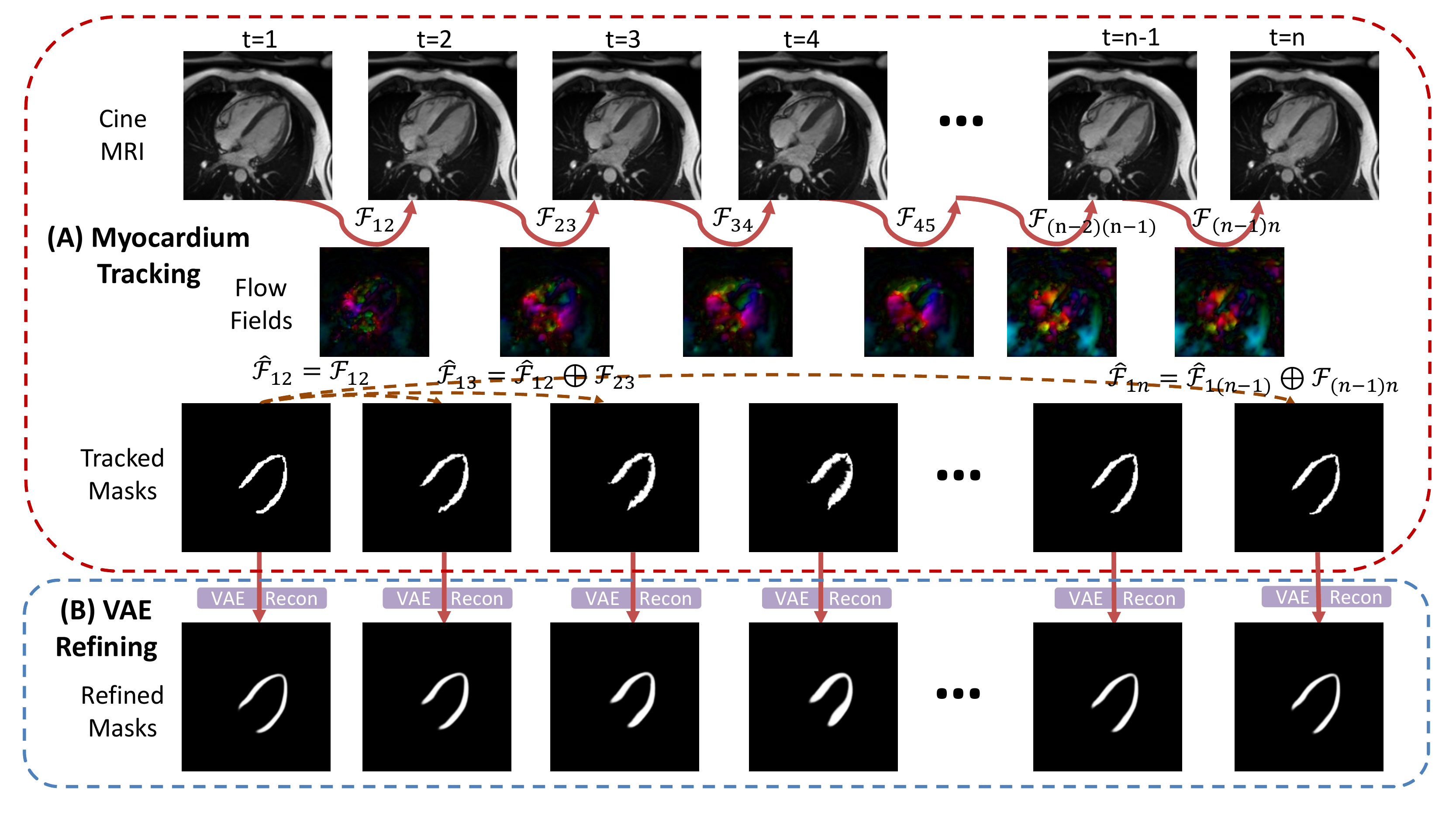}
\caption{Cine MRI myocardium tracking via motion estimation from consecutive frames and myocardium refining via VAE reconstruction. (A) describes the flow field composing procedure and the myocardium tracking based on the composite flow fields. (B) presents the refining of the tracked myocardium using VAE.}
\label{fig:myocardium_tracking}
\end{figure}

We can now perform CMR-FT for cine MRI based on the motion estimation between consecutive frames, as shown in Fig. \ref{fig:myocardium_tracking}(A). We denote the flow field between the (n-1)-th and n-th frame as $\mathcal{F}_{(n-1)n}$. We compute the composite flow field $\hat{\mathcal{F}}_{1n}$ between the first and n-th frame using all intermediate motion fields:
\begin{equation}
    \hat{\mathcal{F}}_{1n} = 
        \begin{cases}
            \mathcal{F}_{12} & n=2\\
            \hat{\mathcal{F}}_{1(n-1)} \bigoplus \mathcal{F}_{(n-1)n} & n>2,
        \end{cases}
\end{equation}
where $\bigoplus$ is a flow composite operator and $\mathcal{F}_{ik} = \mathcal{F}_{ij} \bigoplus \mathcal{F}_{jk} = \mathcal{F}_{ij} \bigotimes \mathcal{F}_{jk} + \mathcal{F}_{jk}$. Note that motion is estimated only between two neighboring frames to avert large feature changes due to image intensity drifting and severe deformation seen in cine MRI. Additionally, the calculation of composite flow always refers to the ED frame because myocardium semantic information (\textit{i.e.}, segmentation mask) is usually given at the ED frame, either by cardiologists' annotation or computer-aided algorithms. Without the loss of generality, we assume the first frame is ED. The myocardium semantic information can thus be obtained by warping the first frame with the composite flow field $\hat{\mathcal{F}}_{1n}$.

We take a refinement step to compensate for potential accumulation error through tracking. Specifically, after the warping to the n-th frame $\mathcal{I}^{'}_{n} = \hat{\mathcal{F}}_{1n} \bigotimes \mathcal{I}_{1}$, motion  $\mathcal{F}_{n}^{\delta}$ is estimated between $\mathcal{I}^{'}_{n}$ and $\mathcal{I}_{n}$. The final compensated motion between the first and the n-th frame is composed as $\hat{\mathcal{F}}_{1n}^{*} = \hat{\mathcal{F}}_{1n} \bigoplus \mathcal{F}_{n}^{\delta}$.

\subsection{Anatomy-Aware Motion Estimation}
\textbf{Shape prior via VAE:} The tracked myocardium using the above pipeline achieves fairly promising results. However, as the baseline model is mainly based on image intensity difference, the estimated motion can thus be severely affected by disturbances such as intensity-similar anatomies (\textit{e.g.}, papillary muscle) and noises, which leads to tracked myocardium with unrealistic anatomy (``Tracked Masks" in Fig.~\ref{fig:myocardium_tracking}(A)). 

To solve this problem, we utilize convolutional VAE \cite{pu2016variational,hou2017deep,higgins2017beta} to encode myocardium anatomy and enforce myocardium shape constraints (anatomy-awareness) in the motion estimation. Using available myocardium annotations, we train the VAE model to take the myocardium mask as input and reconstruct it. In addition to the reconstruction loss, the Kullback-Leibler Divergence (KLD) loss is used to enforce the latent space to conform to a standard normal distribution. Thus, the encoder's outputs are a mean vector and a standard deviation vector. During the training, we sample from this distribution based on the mean and the standard deviation and reconstruct myocardium shape from the sample via the decoder. We used the trained VAE model (both the encoder and the decoder) to correct unrealistic myocardium masks. Specifically, the mean latent variable representing expected myocardium manifold in the latent space given the input mask is used without sampling. The decoder can decode such a latent variable into a realistic shape. Fig.~\ref{fig:myocardium_tracking}(B) shows the reconstructed myocardium using the trained VAE model. 

\textbf{Weakly-supervised motion estimation using shape prior:} For every cine MRI in the training dataset, we first use the baseline model to feature-track the myocardium to obtain coarse myocardium results for every image. The tracked myocardium through time, with possible unrealistic shapes, are then corrected by the VAE model, which are further used as anatomy constraints to improve the motion model. In this way, the motion estimation can mitigate the disturbances in images. Specifically, each input image pair ($\mathcal{I}_{1}$, $\mathcal{I}_{2}$) now has its corresponding corrected myocardium ($\mathcal{M}_{1}$, $\mathcal{M}_{2}$). We apply the flow fields $\mathcal{F}_{12}$ and $\mathcal{F}_{21}$ to their corresponding masks and obtain warped masks $\mathcal{M}_{1}^{'} = \mathcal{F}_{12}\bigotimes\mathcal{M}_{1}$ and $\mathcal{M}_{2}^{'} = \mathcal{F}_{21}\bigotimes\mathcal{M}_{2}$. We expect that a plausible flow field will preserve the anatomy after warping and therefore propose the anatomy loss function $\mathcal{L}_{anat}^{M}= \vert \mathcal{M}_{1} - \mathcal{M}_{2}^{'} \vert + \vert \mathcal{M}_{2} - \mathcal{M}_{1}^{'} \vert$.

Furthermore, we apply the VAE model to the warped masks ($\mathcal{M}_{1}^{'}$, $\mathcal{M}_{2}^{'}$) and obtain their reconstructed masks ($\mathcal{M}_{1}^{recon}$, $\mathcal{M}_{2}^{recon}$). We enforce the warped masks to be close to their VAE reconstructions to constrain the motion estimation model further using the reconstruction loss $
    \mathcal{L}_{recon}^{M}= \vert \mathcal{M}_{1}^{'} - \mathcal{M}_{1}^{recon} \vert + \vert \mathcal{M}_{2}^{'} - \mathcal{M}_{2}^{recon} \vert$.
We define the anatomy-aware motion estimation loss as:
\begin{equation}
    \mathcal{L} = \mathcal{L}_{cons} + \lambda_{H}\mathcal{L}_{H} +  \lambda_{anat}\mathcal{L}_{anat}^{M} + \lambda_{recon}\mathcal{L}_{recon}^{M},
\end{equation}
where $\lambda_{H}$, $\lambda_{anat}$, and $\lambda_{recon}$ are weights, and we denote this model as AATracker. Fig. \ref{fig:motion_estimation} presents examples for the aforementioned images, flow fields, and masks. After refining the baseline model, we then apply the AATracker to the pipeline in Fig. \ref{fig:myocardium_tracking}(A) to track myocardium. Since only the myocardium in the first frame needs annotation, and the rest are tracked and then corrected by VAE, the whole process is weakly-supervised. It is worth pointing out that ED frame annotation is available in clinical setup for CMR-FT application. Also note that the AATracker directly estimates cardiac motions that preserve the underlying anatomy. The anatomy-aware learning module (Fig. \ref{fig:motion_estimation}B) is only performed during the training stage to infuse the anatomical knowledge into the motion estimation network.

\section{Experiments and Results}
\subsection{Implementation Details}
We benchmark the proposed method on 1,137 2-chamber and 1,111 4-chamber cine MRI from Kaggle. Each cine is from one patient with 30 frames. Data is randomly split for training, testing and validation. The VAE model is trained on fully-annotated 100 cine (3000 frames). The motion estimation refinement is trained on another 300 cine with the first frame annotated, and is tested on another 45 fully-annotated cine (1,350 frames). The remaining data, without any annotation, are used for the unsupervised baseline model training, hyper parameters tuning and model selection. All images and annotated myocardium masks are rescaled to the same resolution and cropped into $192\times192$. The images are normalized into zero mean and unit standard deviation. Since no ground truth motion field is available, we evaluate the motion estimation based on the tracked myocardium with three commonly used metrics. The Dice similarity coefficient (DSC) measures the overlapping regions. The Hausdorff distance (HD) calculates the maximum distance between two boundaries, while the average symmetric surface distance (ASSD) calculates the average distance between two boundaries.

The implementation includes three aspects: baseline motion estimation, VAE, and AATracker. The baseline model is trained with $\lambda_{H}$ as 0.02. We train the VAE model with extensive data augmentation, including vertical and horizontal flipping, and multi-angle rotation, and set the latent space as a 32-d representation. In the AATracker training, $\lambda_{H}$, $\lambda_{anat}$, and $\lambda_{recon}$ are set as 0.04, 6.0, 1.2, respectively.  We compare the AATracker with the baseline model and the two shape-constrained models employing either anatomy loss or reconstruction loss. We also compared to two conventional registration methods, multi-scale free form deformation (FFD) \cite{modat2010fast,joshi2001multi} and multi-scale diffeomorphic demons \cite{vercauteren2009diffeomorphic}, that have been previously used for medical image motion estimation \cite{haskins2020deep}. All models are trained and tested on a standard workstation equipped with Intel Xeon Bronze 3106 CPUs and a Nvidia Titan XP GPU.

\subsection{Results and Discussions}
Table \ref{Tab:ModelComparision} presents the quantitative results over compared methods. Both anatomy loss and reconstruction loss can boost the performance of the baseline, while the effect of anatomy loss is more noticeable. The anatomy-aware model AATracker with both losses attains the best performance. Compared with the baseline model, AATracker reduces HD by 12.3\% and 14.9\% on 2- and 4-chamber cine MRI, respectively. The accuracy improvements of utilizing deep-learning-based methods over conventional methods are consistent with existing studies \cite{balakrishnan2018unsupervised,mansilla2020learning}. Besides, the AATracker takes much less time (${\sim}1.5$s) 
on average to estimate the motion than FFD (${\sim}46.1$s) and diffeomorphic demons (${\sim}25.2$s) for a cine MRI of typical size. 


\begin{table}[bt]
\caption{Comparison of FFD, diffeomorphic demons, Baseline model, baseline with anatomy loss (Baseline+anat), baseline with reconstruction loss (Baseline+recon) and the AATracker. Dice Similarity Coefficient (DSC), Hausdorff Distance (HD), and Average Symmetric Surface Distance (ASSD) on 2- and 4-chamber cine MRI are shown in mean(std).}
\label{Tab:ModelComparision}
\begin{tabular}{C{1.6cm}C{3.6cm}C{2.2cm}C{2.2cm}C{2.2cm}}
\hline
Long-Axis &  Method & DSC & HD (mm) & ASSD (mm) \\
\hline
\multirow{5}{*}{2-chamber} &  FFD                 & 0.768 (0.054) & 6.242 (1.556) & 1.480 (0.310) \\
                           & Diffeomorphic Demons & 0.791 (0.051) & 6.265 (1.313) & 1.370 (0.238) \\
                           &  Baseline            & 0.834 (0.039) & 6.389 (1.444) & 1.203 (0.216) \\
                           & Baseline+anat        & 0.835 (0.048) & 5.659 (1.285) & 1.158 (0.209) \\
                           &  Baseline+recon      & 0.835 (0.038) & 6.163 (1.507) & 1.190 (0.224) \\
                           &  AATracker           & \textbf{0.836} (0.048) & \textbf{5.604} (1.252) & \textbf{1.154} (0.206) \\
\hline
\multirow{5}{*}{4-chamber} &  FFD                 & 0.803 (0.039) & 6.067 (1.339) & 1.306 (0.238) \\
                           & Diffeomorphic Demons & 0.813 (0.037) & 6.936 (1.583) & 1.274 (0.213) \\
                           &  Baseline            & 0.861 (0.026) & 6.228 (1.607) & 1.062 (0.207) \\
                           &  Baseline+anat       & 0.864 (0.026) & 5.328 (1.185) & 1.007 (0.163) \\
                           &  Baseline+recon      & \textbf{0.865} (0.025) & 5.936 (1.615) & 1.026 (0.197) \\
                           &  AATracker           & 0.864 (0.028) & \textbf{5.303} (1.171) & \textbf{0.998} (0.160) \\
\hline                           
\end{tabular}
\end{table}

\begin{figure}
\includegraphics[width=\textwidth]{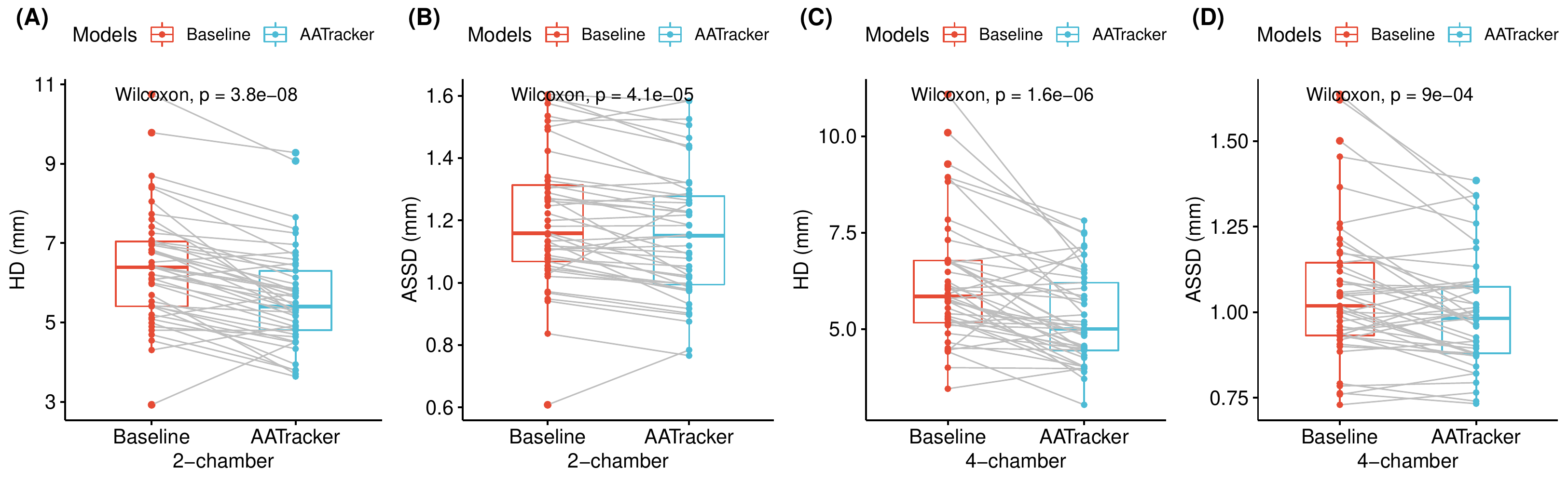}
\caption{Patient-wise comparison between the baseline model and AATracker on 2-chamber (A-B) and 4-chamber (C-D) cine MRI.}
\label{fig:patient_wise_cmp}
\end{figure}

\begin{figure}[bt]
\includegraphics[width=\textwidth]{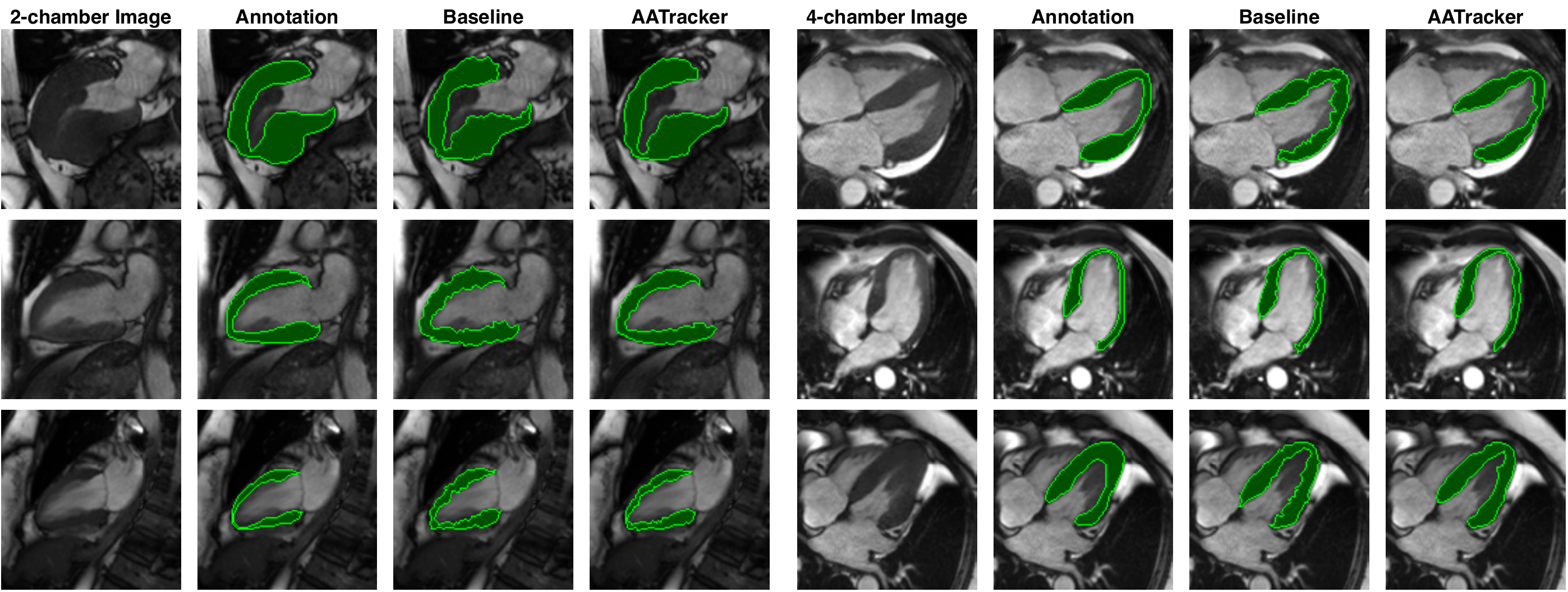}
\caption{Examples of myocardium annotation, baseline, and AATracker results.}
\label{fig:myocardium_mask_cmp}
\end{figure}

Fig. \ref{fig:patient_wise_cmp} shows a patient-wise comparison between the baseline model and AATracker. On both 2- and 4-chamber evaluation, the p-values from the Wilcoxon signed-rank test are significant for the two boundary-based metrics HD and ASSD. This result demonstrates that AATracker consistently improves the myocardium tracking results. Fig. \ref{fig:myocardium_mask_cmp} shows examples of tracked myocardium. The results of AATracker are visually more similar to the annotations. Most importantly, the anatomy-aware myocardium are more anatomically reasonable with smoother boundaries, demonstrating the effectiveness of the shape constraints. 

Both Fig.~\ref{fig:patient_wise_cmp} and Fig.~\ref{fig:myocardium_mask_cmp} demonstrate the improved performance of the AATracker over baseline. These results indicate that AATracker preserves the anatomy structure during tracking. Arguably, DSC improvement is subtle after the motion estimation refinement. The main reason is that the myocardium boundary only accounts for a tiny part in the myocardium, and the refinement works mainly on the myocardium boundary without substantially affecting the overall myocardium shape. The reduction in ASSD is not as significant as HD, likely because ASSD is the average distance considering all boundary pixels while HD measures the worst error distance. 

\section{Conclusion}
We present an end-to-end framework incorporating the anatomy prior to training for the awareness of anatomy in cardiac motion estimation. To our best knowledge, this is the first work that introduces shape constraints into the myocardium feature tracking via weak supervision. The proposed anatomy-aware method achieves consistent improvements over the baseline deep learning methods and two conventional methods. This study provides a sound basis for further cardiac function assessment, such as strain analysis. 

\bibliographystyle{splncs04}
\bibliography{myrefs}

\end{document}